\documentstyle[11pt,revpasp,epsf]{article}

\begin{document}

\title{Evolution of Shape in the Field}

\author{C.H.\,Heller}
\affil{Universit\"ats-Sternwarte, Geismarlandstra\ss e 11,
       37083 G\"ottingen, Germany (cheller@uni-sw.gwdg.de)}

\begin{abstract}

The evolution in shape of an isolated density enhancement in
the early universe is studied through numerical simulations.
The formation scenarios of a cold dissipationless collapse
and that of a slow accumulation of gas in a dark matter halo
are examined.
In the later case the conversion of gas to stars and the
accompanying energy feedback from stellar winds and SNII are
taken into account.  

It is found that with the more realistic initial conditions
the radial orbit instability (ROI) is able to operate at
higher values of the virial coefficient than
previously believed.  In the cold dissipationless collapse the
system becomes fully axisymmetric on a timescale that is less
than one collapse time.  A prolate or oblate figure is obtained
for low and high rotations, respectively.  The inclusion of gas and
star formation leads to halos that are more triaxial, while the
stellar systems that form in them are oblate, independent of the
initial rotation.

\end{abstract}
 
\section{Introduction} 

The transformation of an isolated  primordial gas cloud into a galaxy
may proceed through either a short phase of cold dissipationless
collapse or a longer period of gas accumulation.
A dissipationless collapse requires an early high efficiency ``burst''
of star formation which converts most of the gas to stars.  In contrast
the energy feedback from stellar winds and supernovae may produce a
self-regulated slow accumulation of gas and conversion into stars.

A sufficiently cold initially spherical system was found by Merritt \& Aguilar
(1985) not to retain its spherical shape during collapse.  The symmetry
breaking mechanism is a still not fully understood form of the radial orbit
instability (ROI).  A further work by Aguilar \& Merritt (1990)
showed that the final shape may be either prolate or oblate depending
on the initial amount of rotation.  Other studies have examined the role
played by a variety of different parameters (Cannizzo \& Hollister 1992;
Udry 1993; Hozumi, Fujiwara, \& Kan-ya  1996; Theis \& Spurzem 1999).

In this study the shape evolution of a $2\sigma$ density enhancement in
an Einstein-de~Sitter universe is followed through N-body/SPH simulations.
In the dissipative scenario, the conversion of gas to stars, along with
the accompanying mass and energy feedback is included in the model.
The evolution in the shape of the dark matter halo and of the stellar
component that forms in it will be examined.

\section{Dissipationless Evolution}

\subsection{Methods}

\subsubsection{N-body Code}

The numerical simulations are performed with an updated version
of the N-body algorithm described in Heller (1995) and
Heller \& Shlosman (1994). The integrator used is a second-order
Runge-Kutta, which also provides an estimate of the required time-step for
each particle (Navarro \& White 1993).  A hierarchical set of time bins
is employed,
an important consideration given the large range of dynamical timescales
present in the models.  The gravitational forces are computed using a
surface harmonic method combined with a logarithmic radial grid 
(Sellwood 1997).  For this study terms up to $l=6$ are retained
in the expansion.

\subsubsection{Shape Determination}

To determine the intrinsic shape of the particle distribution we start
by removing any residual net velocity from
the system, followed by rejection of any unbound particles, and iterating
this procedure as required (Aguilar \& Merritt 1990).  Next
we locate the density center as defined by Casertano \& Hut (1985), 
\begin{equation}
{\boldmath r}_{\rm d} = \frac{\sum \rho_i {\boldmath r}_i}{\sum \rho_i},
\end{equation}
where $\rho_i$ is the local mass density at the position of each
particle.  This local density is evaluated using a kernel based
expectation value as used in SPH methods (Monaghan 1992),
\begin{equation}
\rho_i = \sum m_j W(r_{ij},h_i),
\end{equation}
where $2h_i$ is the radius of a sphere centered on particle $i$ which
contains a fixed number of neighboring particles, taken here as 
${\cal N}=96$.  The kernel $W$ is given by a normalized spline
function (Hernquist \& Katz 1989).

In the reference frame defined by the so determined velocity centroid
and density center, we compute the eigenvalues of the moment of
inertia tensor.  From these we may determine the axes $a>b>c$ of a uniform
spheroid with the same eigenvalues.  The ratios of these
axes may then be used to characterize the shape of the system.
Typically the ratio of the shortest to longest axis, $c/a$, is
given for different fractions of the particles sorted by
binding energy.

\subsubsection{Initial Conditions}

The initial conditions are that of a spherically symmetric density
enhancement in an Einstein-de~Sitter ($\Omega=1$) universe
as given by Thoul \& Weinberg (1995). The initial density
profile is that of the average density around a $2\sigma$ peak in a
Gaussian random density field with power spectrum 
$P(k) = A k^{-2} \exp\left(-k^2 R_f^2\right)$. 
The particles are initially moving outward with the Hubble flow, with 
consecutive shells of mass stopping, turning around and falling back
inward.  The model is conveniently defined by the 
filter mass $M_f$, which defines the mass contained within a sphere
of radius $2\,R_f$, and a collapse redshift $z_c$, which defines the
redshift at which this shell reaches the center.  In
addition to these the initial conditions are set by a 
virial coefficient $q=2T_{\rm rnd}/\left|U\right|$, where $T_{\rm rnd}$ is
the kinetic energy in random motions and $U$ is the potential energy,
and a spin parameter $\lambda=J\left|E\right|^{1/2}G^{-1}M^{-5/2}$,
where $J$, $E$, and $M$ are respectively the total angular momentum,
energy, and mass of the system. 

As a standard model we adopt a non-rotating model with $M_f=1$, $z_c=2$,
$q=0.05$, and $\lambda=0.0$.  Following Thoul \& Weinberg (1995)
all models are started at an initial redshift of $z_i=36$ to ensure they
begin in the linear regime.  For reference, with mass and distance units of,
respectively, $10^{10}\,{\rm M}_\odot$ and 1\,kpc, and a Hubble constant
of $75\,{\rm km}\,{\rm  s}^{-1}\,{\rm Mpc}^{-1}$, the model would have a
filter radius $R_f=3.2$\,kpc and a circular velocity of $v_c=54$\,km/s.

\subsubsection{Tests}

Many tests were performed to check the sensitivity of the results
to the algorithm and its parameters. Tests with half and twice
the number of particles give essentially identical results.  Changing the
grid resolution only effects the very inner region, while making
no noticeable difference to the overall evolution.
A typical run conserves energy to about one percent.

As a check of the gravity solver the standard model was also run using a
BH-tree (Barnes \& Hut 1987) with quadrapole terms 
(Hernquist 1987) and the Grape-3Af special purpose hardware
(Sugimoto et al. 1990).  The 
integration was also checked against that of a standard leap-frog.  
The results are shown in Fig.~1 
\begin{figure}[t]
{\centering\leavevmode\epsfxsize=0.9\columnwidth
 \epsfbox[70 200 480 480]{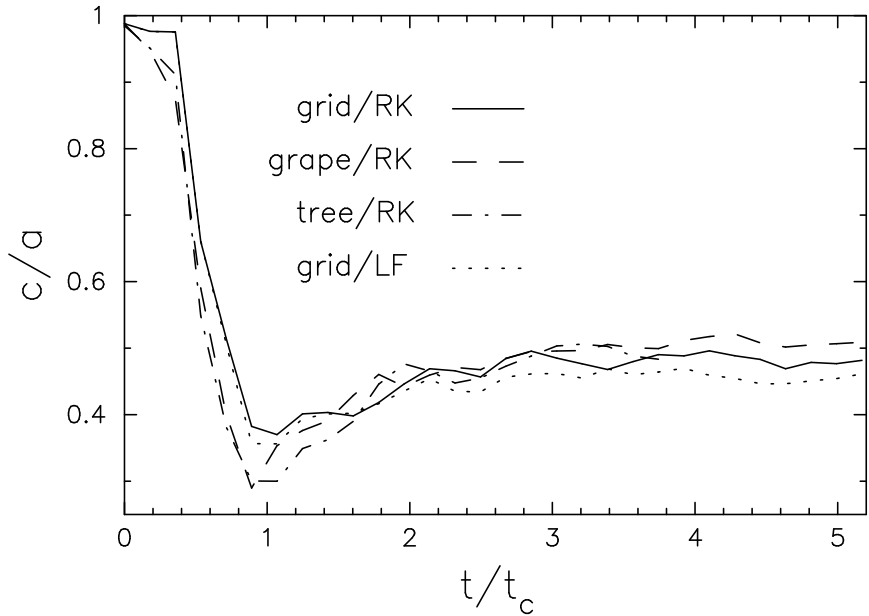}}
\caption{Comparison of results for different combinations of
integrator and gravity solver.  Shown is the axis ratio $c/a$ of
the most bound 60\% of the particles vs. time in units of the collapse
time.  The integrators used are a second-order Runge-Kutta (RK)
and a standard leap-frog (LF).  The gravity solvers are a spherical
harmonic grid, a BH-tree, and a Grape-3Af.} 
\end{figure}
where the axis ratio $c/a$ is given for the most
bound 60\% of the particles as a function of time
in units of the collapse time $t_{\rm c}$.  
It should be noted that
the evolution is independent of the both the mass of the model and
the collapse redshift when given in these units. 
The two integrators
give similar results, though the leap-frog shows perhaps
slightly less fluctuations.  The tree and Grape results also match each
other closely, but compared to the grid produce a greater maximum elongation.
The difference is likely due to the higher central resolution provided by
the grid.

\subsection{Results}

\subsubsection{Standard Model}

The evolution of the shape for the standard model (collisionless) 
is shown in Fig.~2,
\begin{figure}[t]
{\centering\leavevmode\epsfxsize=0.9\columnwidth
 \epsfbox[70 200 480 480]{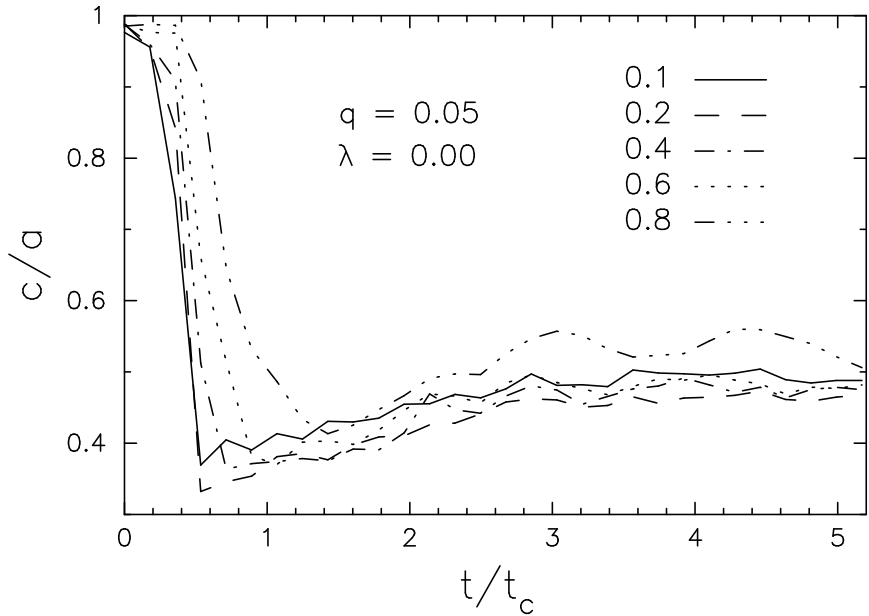}}
\caption{Evolution of the shape of the standard model.  Shown
is the axis ratio $c/a$ for different fractions of the most
bound particles as a function of time in units of
the collapse time.
The collapse time of the model is defined as the time at which
the shell at radius $2\,R_f$ reaches the center for cold initial
conditions.} 
\end{figure}
where the axis ratio $c/a$ is given for different fractions of
the most bound particles, $f_b$, as a function of time.
The time at which maximum
elongation is obtained increases with the binding fraction, 
in support of the previous finding that the instability does
not operate until around the time of maximum collapse
(Hozumi, Fujiwara, \& Kan-ya  1996; Theis \& Spurzem 1999).
In addition to this the maximum value obtained decreases with
the binding fraction.  After the initial collapse, between 
$1-3\,t_{\rm c}$, the elongation becomes slightly less, reaching
an nearly equilibrium value of around $c/a=0.48$ for most of the system.  
The deviations displayed by the
$f_b=0.8$ curve are attributed to both edge effects introduced by
truncating the initial distribution at $r=2\,R_f$ and the longer
relaxation time in the outer parts.
The final relaxed shape is highly prolate, with the ratio of the
two shorter axes $c/b=0.98$.  The corresponding triaxiality
index $\tau\equiv(b-c)/(a-c)$ is around 0.02.  

The surface density profile of the standard model at $t=4\,t_{\rm c}$
is given in Fig.~3.
\begin{figure}
{\centering\leavevmode\epsfxsize=0.9\columnwidth
 \epsfbox[70 200 480 480]{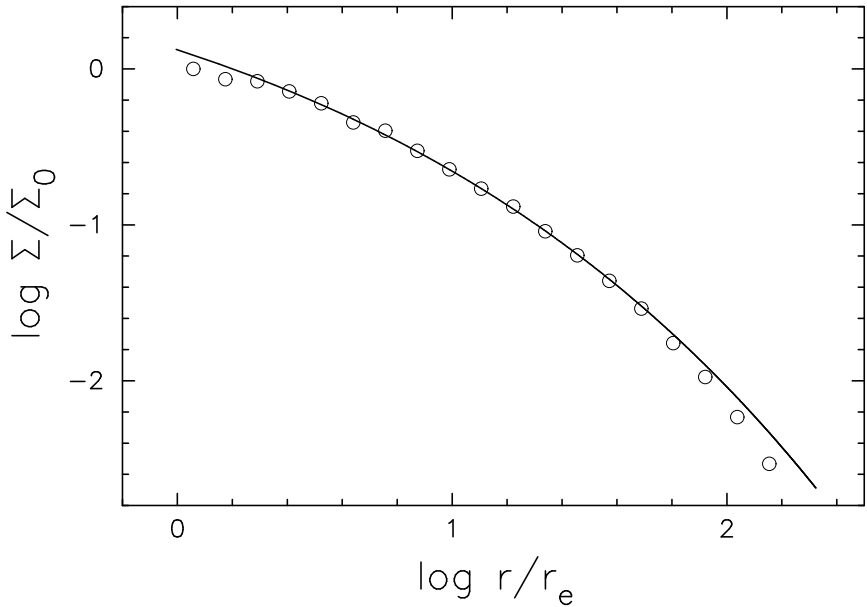}}
\caption{Normalized surface density profile of the relaxed standard
model.  The abscissa is given in units of the effective radius
obtained from the best-fit $r^{1/4}$ law, which is indicated by
the solid line.}
\end{figure}
Shown is the normalized surface density as a function
of radius.  The surface density is measured in logarithmically
spaced elliptical annuli aligned with the principal axes of the inertia
tensor and with the same ellipticity.  Also shown is a fit to the
functional form $\Sigma\propto \left(r/r_e\right)^{1/4}$.  The
radius is given in units of this effective radius $r_e$, which in
terms of the filter radius has a value of about $0.05\,R_f$.  As can
be seen the surface density of the relaxed model is well fit by
an $r^{1/4}$ law over a range in radius of $2-50\,r_e$.  
Inside of this range, numerical resolution has affected
the results, while outside, the system is still not fully relaxed.

\subsubsection{Virial Coefficient}

In Fig.~4 
\begin{figure}
{\centering\leavevmode\epsfxsize=0.9\columnwidth
 \epsfbox[70 200 480 480]{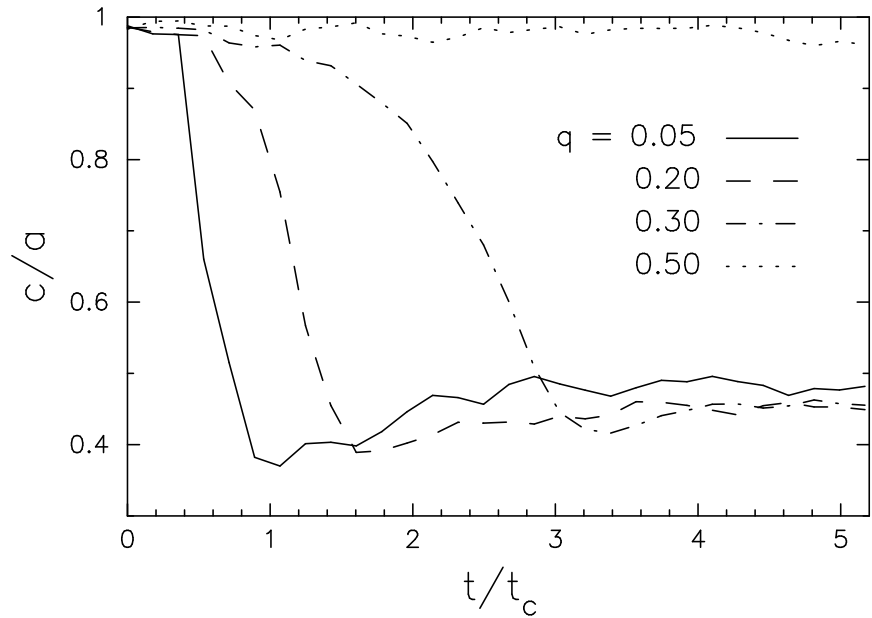}}
\caption{Evolution of the axis ratio of the most bound 60\% of
the particles for different values of the initial virial
coefficient.  The time is given in units of the collapse time.}
\end{figure}
the evolution of the axis ratio $(c/a)_{0.6}$ is given for
different initial values of the virial coefficient.  As the value
of $q$ is increased the maximum elongation attained decreases 
slightly and the time at which it occurs is increasing delayed.  
However, the final shape is rather insensitive to the value of $q$.
This is true even for the relatively hot model with $q=0.3$ 
which reaches the same elongation as the colder models after
about three collapse times. It is also found (not shown) that the
instability is damped to progressively larger radii as $q$ is increased.
Only with a remarkably high value of the
virial coefficient is the ROI fully damped.

\subsubsection{Rotation}

The non-rotating standard model shown in Fig.~2 may be directly compared
with a rotating model ($\lambda=0.07$) in Fig.~5.
\begin{figure}
{\centering\leavevmode\epsfxsize=0.9\columnwidth
 \epsfbox[70 200 480 480]{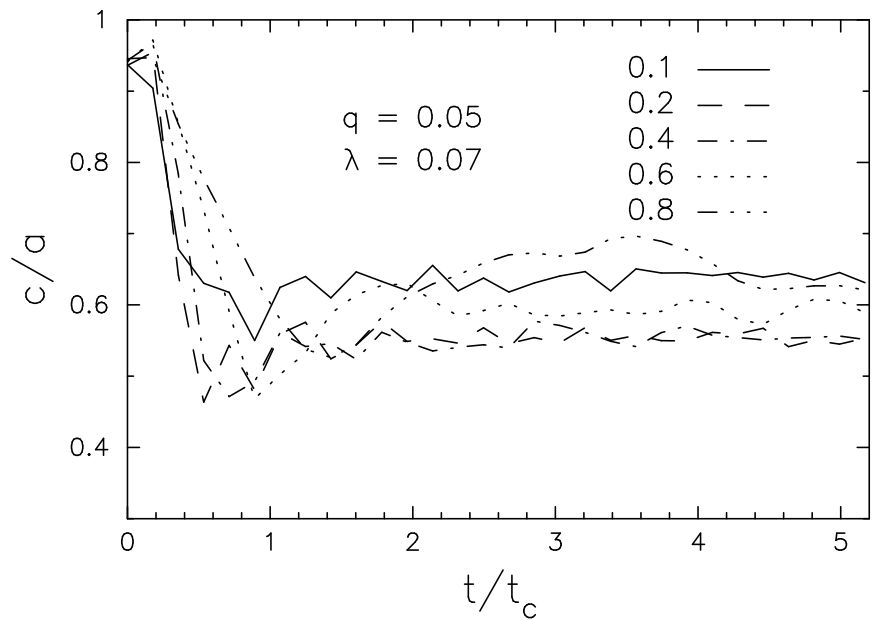}}
\caption{Evolution of the shape of a model with initial
spin parameter $\lambda=0.07$.  Shown is the axis ratio $c/a$ for
different fractions of the most bound particles as a function of time
in units of the collapse time.}
\end{figure}
The rotating model is rounder and the difference between maximum elongation
and the final shape significantly reduced.  There is also a trend in the
outer 50\% of the mass of decreasing elongation with binding fraction. 
The model is also significantly flattened, with $(c/b)_{0.6}=0.6$, 
with again a slight trend of increasing with binding fraction.  The
corresponding triaxiality index is $\tau=0.99$, indicating a
highly oblate figure. While the number of escapers increases with the
introduction of rotation, the value is still rather small, with only
about 2\% of the particles having positive energies at $t=4\,t_{\rm c}$.

\section{Dissipative Component}

\subsection{Methods}

The gas component is simulated using the smooth particle hydrodynamic (SPH)
method, where the continuous fields (e.g. velocity, density, internal energy, etc.)
are evaluated from a set of points or particles which move with the fluid.
The gravitational forces and SPH neighborhoods are computed using the
Grape-3Af special hardware.  The initial model consists of 5000 SPH particles
representing 10\% of the mass, distributed with the same radial profile as
the collisionless particles which represent the dark matter.

\subsubsection{Heating and Cooling}

Sources of heating and cooling in the energy equation include
adiabatic, viscous, and radiative.  The net rate to the specific
internal energy,  along with the ionization fractions 
(H, H+, He, He+, He++, e) and mean molecular weight are computed
as a function of density and temperature for an assumed
optically thin primordial composition gas (Katz et al. 1996).

\subsubsection{Star Formation}

Star formation takes place in regions which are not expanding, are 
at least
moderately self-gravitating in the background of stars and dark matter,
and where the collapse would continue unhindered if given sufficient
numerical resolution.  Specifically, the criteria are given by,
\begin{enumerate}
\item $\nabla\!\cdot\!\vec{v} \le 0$,
\item $\rho_{\rm HI}  > \frac{1}{2}\rho_{\rm vir}$, 
\item $\tau_{\rm cool} \ll \tau_{\rm dyn} < \tau_{\rm sound}  
      \ \rightarrow \ \rho_{\rm HI} > \rho_{\rm crit}$,
\end{enumerate}
where the virial density 
$\rho_{\rm vir}=\rho_{\rm gas}+\rho_{\rm star}+\rho_{\rm dm}$ includes
the combined mass of gas, stars and dark matter. The critical density is
taken as $7\!\times\!10^{-26}\,{\rm g/cm}^3$.   Gas is converted into
 stars at a rate 
\begin{equation}
\dot{\rho}_{\rm gas} = -\frac{\rho_{\rm HI}}{\tau_{\rm c}}, \label{sfrate}
\end{equation}
which implies a time scale for star formation of
\begin{equation}
\tau_\star = -\tau_{\rm c}\frac{\rho_{\rm gas}}{\rho_{\rm HI}}
\ln\left(1-\eta\right),
\end{equation}
where $\eta$ is a star formation efficiency parameter which controls
the amount of mass converted in each star formation event, and therefore
determines the number of events per time $\tau_\star$ required to obtain
the rate given by Eqn.~\ref{sfrate}. The efficiency and star formation collapse
timescale is taken here as $\eta=0.3$ and $\tau_{\rm c}=10\,\tau_{\rm dyn}$,
respectively.

Metal enriched mass (40\%) is returned to the surrounding gas by the newly 
formed stellar particles.  Also energy from stellar winds and SNII is
injected into the gas, added a little at a time at every timestep
based on a Salpeter IMF and lifetime vs. mass relation.
This feedback energy is injected in both mechanical (1\%) and thermal
form (1-10\%), dependent on the local gas density and metallicity
(Thornton, 1998).

\subsection{Results}

A slow rotating model with $\lambda=0.02$ and fast rotating model
with $\lambda=0.07$ were run.  The other parameters are the same as
the standard model without gas defined in Sec.~2.1.  

Star formation
in both models starts near the center at $t\approx0.3-1.0\,t_{\rm c}$
with 1-3 bursts of 0.5--3.0\,${\rm M}_\odot/{\rm yr}$,
before transitioning to a low continuous rate of around
$0.1-0.2\,{\rm M}_\odot/{\rm yr}$.  In both models almost all of
the star formation occurs within $0.3\,R_f$ of the center. The 
large gas disk ($R\approx1.5\,R_f$) that forms in the high rotation model
remains smooth and axially symmetric, never reaching a high enough surface
density to clump or form spiral features which would promote star formation.

The shape evolution of the halo in the slow rotating model is shown
in Fig.~6.  
\begin{figure}
{\centering\leavevmode\epsfxsize=0.9\columnwidth
 \epsfbox[70 200 480 480]{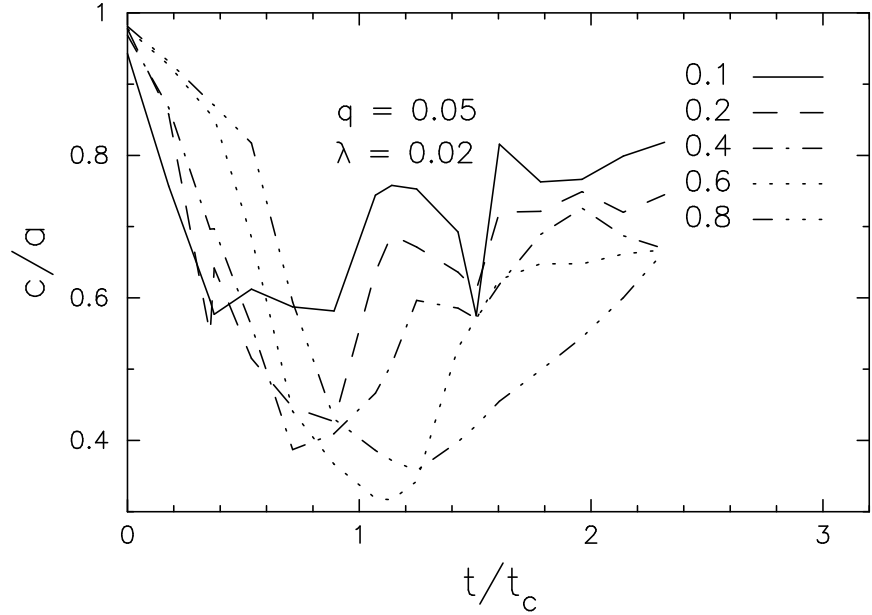}}
\caption{Evolution in shape of a slowly rotating dark matter halo in
a model which includes gas and star formation. The initial spin
parameter is $\lambda=0.02$.  Shown is the axis ratio $c/a$ for
different fractions of particles as a function of time
in units of the collapse time.}
\end{figure}
The maximum elongation of the inner regions has been significantly reduced
by the mass accumulation and ejections at the center.  
The overall shape after $2\,t_{\rm c}$ is rounder than in the non-dissipative
case and with a larger spread in axis ratios, $c/a\approx0.6-0.8$.  
To determine if this spread will tighten up with time, the model
must be continued longer than what is shown here.  The halo is also
less prolate with a triaxiality index $\tau=0.33$.  The stellar
component that has formed at the center of the halo is in contrast
oblate ($\tau=0.90$), with the short axis lying perpendicular
to the halo long axis.

The elongation of the faster rotating halo, shown in Fig.~7,
\begin{figure}
{\centering\leavevmode\epsfxsize=0.9\columnwidth
 \epsfbox[70 200 480 480]{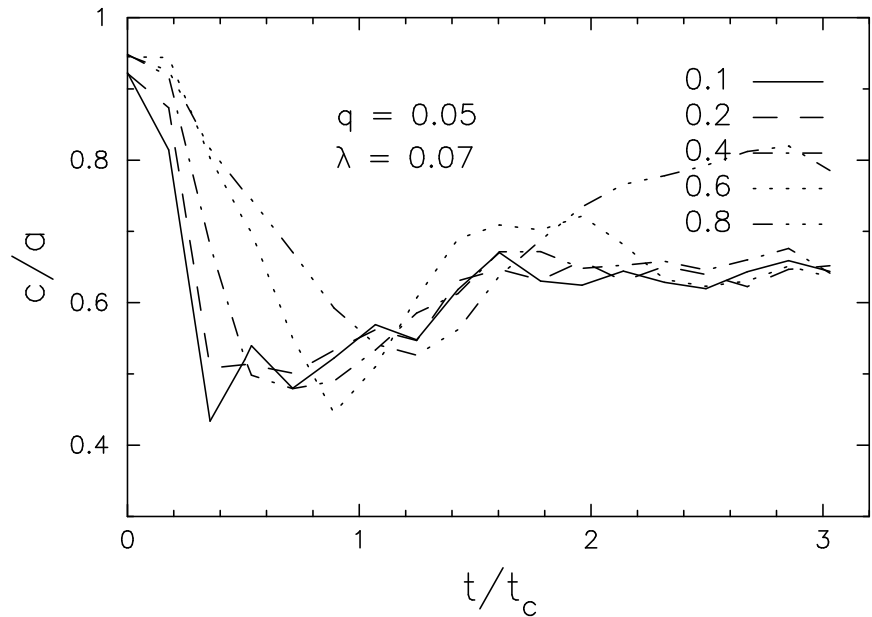}}
\caption{Evolution in shape of a fast rotating dark matter halo in
a model which includes gas and star formation. The initial spin
parameter is $\lambda=0.07$.  Shown is the axis ratio $c/a$ for
different fractions  particles as a function of time
in units of the collapse time.}
\end{figure}
is in contrast less affected by the gas inflow.
This is because the higher rotation has reduced the central concentration
of gas and stars.  The stellar bulge that has formed at the center
of the halo and large gas disk contains only about a third of the mass
as the slow rotating model at any given time. However,
similar to the slow model, the halo has become more triaxial, with
$\tau=0.58$ for $f_b\la0.2$ and $\tau=0.85$ for $f_b\ga0.4$.
The stellar bulge is oblate ($\tau=0.84$) with the short axis aligned
with the short axis of the halo.

\section{Discussion}

Most previous studies of isolated shape evolution have adopted initial
conditions in which the collapse starts from a configuration with no
net radial motions.  This has been justified under the assumption that
it represents a reasonable approximation to a cloud in the early 
universe which as decoupled from the Hubble flow and is just starting
to turn-around after having reached its maximum extent.  However, the
collapse of a density enhancement as described by a typical cosmological
fluctuation spectrum is intrinsically different, in that at no time are
all mass shells at rest with each other.  This property produces a
unique profile of relative collapse times.

In the dissipationless collapse the system reaches full axisymmetry at
around $0.5\,t_{\rm c}$, while the time of maximum elongation
varies with the mass shell from about $0.5-1.0\,t_{\rm c}$.
This is in contrast to the models of Theis \& Spurzem (1999) which
first go through a phase of maximal triaxiality before becoming
axisymmetrized over a period of many dynamical times.
The decrease in elongation which occurs between $0.5-4.0\,t_{\rm c}$
operates on a timescale which is much shorter than that of the two-body
relaxation timescale and may be due to a collective effect.
The phase of violent relaxation is very efficient as
evidenced by the $r^{1/4}$ surface density profile and lack of any
core-halo structure.  The final shape of the system is either prolate
or oblate, respectively, in the cases of low or high rotation, in
agreement with Aguilar \& Merritt (1990).

A difference with previous studies is found in the relatively large values
of the virial coefficient at which the ROI can still operate.  Aguilar \&
Merritt (1990) found at values $q\ga0.1$ the instability was fully damped.
In contrast the dissipationless non-rotating model produced the same overall
final shape at values as high as $q=0.3$.  Such high values
makes the existence of isolated round dark matter halos unlikely.

The presence of gas and star formation produces a more triaxial final
shape.  In terms of elongation, the halo with low rotation shows a greater
response to the dissipative component, particularly the inner regions where 
it is significantly reduced.
In both the low and high rotation models, the stellar component that forms
in the dark matter halo has an oblate shape, with the short axis lying
perpendicular to the halo long axis.  This result is unlikely to be
sensitive to the adopted star formation recipe as long as the potential
is dominated by the dark matter.  For a baryonic dominated potential
the shape of the stellar component that is formed may depend on several
different parameters, such as the star formation feedback efficiency, 
IGM pressure, or the presence of an external UV field.

\acknowledgements
The author wishes to thank J. Sellwood for providing his notes on the
gravity solver, C. Theis for time on the Kiel University Grape, and
I. Shlosman for helpful discussions.  This work has been supported by DFG
grant Fr 325/39-1, 39-2.

\end{document}